\documentclass[%
 reprint,
 superscriptaddress,
nofootinbib,
 amsmath,amssymb,
 aps,
]{revtex4-2}
\usepackage[utf8]{inputenc}
\usepackage[T1]{fontenc}
\usepackage{graphicx} 
\usepackage[svgnames,dvipsnames]{xcolor}
\usepackage{braket}
\usepackage{soul}
\usepackage{amsmath}
\usepackage{bbold}
\usepackage{dcolumn}
\usepackage{bm}
\usepackage{hyperref}
    \hypersetup
    {   
        unicode=true,           
        pdftoolbar=true,        
        pdfmenubar=true,        
        pdffitwindow=false,     
        pdfstartview={FitH},    
        pdfauthor={Edoardo Zavatti, Gabriele Bellomia, Massimo Capone},
        pdftitle={Quantum magic of strongly correlated fermions - The Hubbard dimer},
        pdfkeywords={nonstabilizerness, quantum magic, quantum resources, Hubbard model, strongly correlated electrons, strongly correlated fermions, nongaussianity, entanglement},
        pdfnewwindow=true,      
        colorlinks=true,        
        linkcolor=Maroon,       
        citecolor=Cerulean,     
        filecolor=Maroon,       
        urlcolor=Cerulean       
    }
\usepackage{cleveref}
\crefname{equation}{Eq.\!}{Eqs.\!}
\crefname{figure}{Fig.\!}{Figs.\!}
\crefname{section}{Sec.\!}{Secs.\!}
\usepackage{tikz}
\usepackage{mathtools}
\usepackage{gensymb}
\usepackage{svg}
\usepackage{amssymb}
\usepackage{xfrac}
\usepackage{siunitx}
\usepackage[utf8]{inputenc}
\usepackage[autostyle]{csquotes}

\usepackage{lipsum}

\usepackage{ragged2e}
\makeatletter
\renewcommand{\@makecaption}[2]{%
  \vskip\abovecaptionskip
  \justifying
  \small #1: #2\par
  \vskip\belowcaptionskip}
\makeatother

\newcommand{\LR}{\ensuremath{L\mathcal{R}}}

\usepackage[normalem]{ulem} 

\begin{document}

\title{Quantum magic of strongly correlated fermions -- the Hubbard dimer}
\author{Edoardo Zavatti}
\email{ezavatti@sissa.it}
\affiliation{SISSA, Scuola Internazionale Superiore di Studi Avanzati, via Bonomea 265, 34136 Trieste, Italy}
\author{Gabriele Bellomia}
\email{gabriele.bellomia@tuwien.ac.at}
\affiliation{SISSA, Scuola Internazionale Superiore di Studi Avanzati, via Bonomea 265, 34136 Trieste, Italy}
\affiliation{Institute of Solid State Physics, Technische Universität Wien, 1040 Vienna, Austria}
\author{Massimo Capone}
\affiliation{SISSA, Scuola Internazionale Superiore di Studi Avanzati, via Bonomea 265, 34136 Trieste, Italy}

\begin{abstract}
We study the non-stabilizerness (quantum magic) content of the  Hubbard dimer,
an analytically solvable, yet completely non-trivial, model of strongly correlated fermions. We consider zero- and finite-temperature properties as well as the time evolution after a  quantum quench drives the system out of equilibrium.
We evaluate local and nonlocal non-stabilizerness using both the robustness of magic and the stabilizer Renyi entropy, demonstrating how the latter often fails in detecting the mixed stabilizer states that are typically found in this kind of systems. 
Finally, we compare the non-stabilizerness with other genuine resources of quantum-state complexity, i.e., the fermionic non-Gaussianity and the superselected two-site entanglement. 
Our findings corroborate the role of non-stabilizerness as a fundamental quantum resource, capturing aspects of quantum complexity that elude traditional information-theoretic measures and providing a novel perspective on fermionic systems with tunable interactions.
\end{abstract}

\maketitle


\section{Introduction}
\label{sec: Introduction}

One of the key challenges in achieving quantum advantage lies in understanding how to properly measure the complexity of quantum states, a factor that fundamentally shapes the power of quantum computers and simulators. Historically, entanglement has been regarded as the most prominent manifestation of this complexity and it has long been considered the primary quantum resource \cite{Plenio_EntanglementMeasures,HorodeckiEntanglement,PreskillQC,JozsaEntanglementQC}. However, there are many aspects beyond entanglement that contribute to the intrinsic complexity of a quantum state, and additional notions are required to get a clear assessment of quantum resources. 

A central concept in this search is \emph{non-stabilizerness}, or \emph{magic}, defined as the distance from the convex hull of stabilizer states (stabilizer polytope). 
Since these latter states can be simulated efficiently on classical computers within the Clifford formalism \cite{gottesmanstabilizercodes,gottesmanfaultolerant,gottesmantheorem,nielsenandchuang},  magic quantifies the difficulty to simulate classically a state.
Furthermore, stabilizer states can be implemented fault-tolerantly in several 
quantum architectures \cite{gottesmanerrorcorrection,zeng2011transversality,resourcetheorystabilizer,ROMproperties},
and the presence of non-Clifford resources 
has been shown to be essential for universal quantum computation \cite{kitaevuniversalQC}, and even connected to direct measures of complexity in concrete quantum
algorithms, like Shor's factorization \cite{Secli_Magic_Shor}. 
Hence non-stabilizerness serves as a meaningful measure of the 
intrinsic quantum complexity of a state. 

Non-stabilizerness has gained attention in the field of quantum many-body systems \cite{manybodyquantummagic,aliosciaising,poetrispinchains,SreFazio,SreRydberg,SreKivelson,SreViscardi,SreGravity,PoetriGauge,SreMpsHaug,SreMpsLami,SreMpsPoetri, mariofermionicgaussian,magicmolecules,timsina2025robustness}, where it also provides an additional lens for studying quantum dynamics and phase transitions.

In this work, we focus on the quantum magic of interacting fermionic systems. This investigation inevitably faces with extra layers of complexity.
In particular, already non-interacting (Gaussian) fermionic states display entanglement even if they cannot generate universal quantum complexity \cite{MarianNG,LamiNG,LumiaNG,LyuNG,Mele_NonfreenessLearning,Sierant2025_FAF,MatchgatesNonGaussianity,MatchgatesNonGaussianity2}. As a consequence, fermonic Gaussian circuits describing their evolution -- also known as matchgates -- can be efficiently simulated with classical algorithms \cite{MatchgatesSimulation1,MatchgatesSimulation2,PollmannMatchgates}.

Despite this non-trivial character, a Gaussian state describes in principle non-interacting fermions. For interacting fermions, an approximate effective Gaussian description is obtained via the Hartree-Fock scheme, which is however expected to break down when the correlations between fermions dominate. This \emph{strongly correlated} regime leads to a rich and celebrated phenomenology, ranging from high-temperature superconductivity to a variety of unconventional quantum states. 

In this context, \emph{non-Gaussianity}, also called \emph{non-freeness} 
\cite{gottlieb0,gottlieb1,gottlieb2,vollhardtnonfreeness,Held_nonfreeness,schillingnaturalorbitals}, 
has emerged as a fundamental tool to quantify deviations from the set of free-fermion states, and has proven to be necessary to achieve relevant tasks in quantum information processing \cite{MatchgatesNonGaussianity,MatchgatesNonGaussianity2,Paris_nongaussianity,gaussian1,gaussian2,gaussian3,nogogaussian,coherentstates,coherentstates2}. Moreover, in close analogy with magic-state injection in stabilizer circuits, suitable non-Gaussian resources are sufficient to recover universal features in random matchgate circuits \cite{QuantumResources_RMP,Lami2025_nongauss_doping}.
Finally, most previous studies of non-stabilizerness for fermions are limited to pure states, even if a proper treatment of mixed states  is however necessary to connect with experimental realizations, both in cold-atom and in solid-state platforms.

In this work we address these issues by focusing on a simple, yet non-trivial, strongly interacting fermionic model, namely a two-site Hubbard model, or Hubbard dimer, where we can perform simple exact calculations of different estimates of magic and compare them with non-Gaussianity for both pure and mixed states. 

The Hubbard model is a cornerstone of condensed matter physics, that provides a conceptual framework for strong correlation physics and plays a central role in the study of high-$T_\mathrm{c}$ superconductivity and the Mott metal–insulator transition. 
Within Hubbard-like models, considerable effort has been devoted to characterizing entanglement
\cite{SchillingParadox,SchillingSSRs,SchillingInterorbital,SchillingReview,bellomia2024quasilocal,BippusTwoSiteEntanglement,Held_QFI_pseudodap,bellomia2024quantum}, 
given its well-established relevance for quantum information processing tasks. At the same time, it has been recognized that entanglement between particles can be entirely absent at the local level, as single-site non-Gaussianity is fully determined by classical inter-flavour correlations \cite{Bellomia_intracorr,Zavatti2025}. These findings indicate that entanglement alone does not exhaust the structure of correlations and complexity in interacting fermionic systems. Yet, a systematic investigation of non-stabilizerness in Hubbard systems is still missing. 

The half-filled Hubbard dimer provides an ideal playground for exact calculations at both zero and finite temperature, where mixed states naturally arise. Despite the small size, two-site models feature a non-trivial behavior that mirrors that of large systems and it is often used as a testbed of approximate theories for the Hubbard \cite{Carrascal_2015,Giarrusso} and related models \cite{Avella,PhysRevB.65.104409,Sacchetti_2006}. The simplicity of the model allows us to address also the non-equilibrium dynamics following a quantum quench, explicitly demonstrating how properly accounting for mixed states  becomes crucial in the presence of decoherence. 

By comparing different estimators of quantum magic, we uncover features that are not reflected by entanglement and non-Gaussianity, establishing non-stabilizerness as a distinct resource for characterizing fermionic mixed states. 
Our results establish a direct connection between strong \emph{nonlocal magic}
\cite{gravitationalbackreaction,nonlocalmagic,nonlocalmagic_gaussian1,nonlocalmagic_gaussian2} and the onset
of localized fermionic states, and reveal that nonlocal magic can disappear when 
temperature and decoherence are taken into account. 

Overall, we identify the robustness of magic \cite{ROMproperties} as the most trustworthy quantity to quantify non-stabilizerness in general fermionic states, while the widely used stabilizer Rényi entropy \cite{SREalioscia} turns out to provide unfaithful results for mixed states. This is in close analogy to the role of the von Neumann entropy in entanglement theory: while it is a faithful measure of bipartite entanglement for pure states, it loses this property when applied to mixed states. In the latter case, no simple entropy-based expression generally exists, and the quantification of entanglement typically requires difficult optimization procedures over the set of separable states \cite{HorodeckiEntanglement}.

Our work demonstrates that quantum magic can be successfully studied in strongly interacting fermionic systems, and it offers a different quantum resource with 
respect to non-Gaussianity. Our simple analysis paves the way for investigations for larger systems in different dimensionalities, including applications exploiting accurate numerical approaches to study the celebrated two-dimensional Hubbard model as well as models of different quantum materials.


This article is structured as follows. After reviewing the fundamentals of non-stabilizerness in \cref{sec: Preliminaries}, and briefly describing the model and its exact solution in \cref{sec: Model}, we show our results at zero temperature, in \cref{sec: ZeroT}, as well as at finite temperature in \cref{sec: FiniteT}. In section \cref{sec: Dynamics} we focus on the dynamics of non-stabilizerness after a quantum quench in the Coulomb repulsion, and finally we devote \cref{sec: Conclusions} for final discussions and outlooks.  


\section{Preliminaries}
\label{sec: Preliminaries}

In this section we briefly review the non-stabilizerness measures that we use throughout the text, starting from a short introduction to the Majorana-Clifford group and stabilizer states, in \cref{sec: Clifford_group}, which are the building blocks for the theory of magic. We then introduce the robustness of magic in \cref{sec: ROM}, a magic monotone well suited for mixed states, and finally present the Stabilizer Renyi entropies in \cref{sec: SRE}, highlighting their advantages and limitations when applied to mixed states.

\subsection{Majorana-Clifford group and stabilizer states}
\label{sec: Clifford_group}

We consider a system of $N$ fermionic spin-orbitals (qubits) $\hat{c}_i$, $\hat{c}^\dagger_i$, which satisfy anti-commutation relations $\{\hat{c}_i, \hat{c}^\dagger_j\}=\delta_{ij}$ and span a Hilbert space of dimension $d=2^N$. Notice that in this context the label $i$ includes all possible indexes, such as lattice site, spin and atomic orbital. We denote the local Pauli operators by $\{ \mathbb{\hat{1}}_i, \hat{X}_i, \hat{Y}_i, \hat{Z}_i \}$ and define Majorana operators by the standard Jordan-Wigner mapping, as follows:
\begin{align*}
    \hat{\gamma}_{2i-1} & =  \hat{Z}_1 \otimes ... \otimes \hat{Z}_{i-1} \otimes \hat{X}_i \otimes \mathbb{\hat{1}}_{i+1} \otimes ... \otimes \mathbb{\hat{1}}_N \\
    \hat{\gamma}_{2i} & = \hat{Z}_1 \otimes ... \otimes \hat{Z}_{i-1} \otimes \hat{Y}_i \otimes \mathbb{\hat{1}}_{i+1} \otimes ... \otimes \mathbb{\hat{1}}_N,
\end{align*}
where $i=1,...,N$. Notice that these $2N$ operators are hermitian, and satisfy fermionic anti-commutation relations $\{\hat{\gamma}_i, \hat{\gamma}_j\}=2\delta_{ij}$. Furthermore, they can be rewritten in terms of fermionic creation and annihilation operators as $\hat{\gamma}_{2i-1}=\hat{c}_i+\hat{c}_i^\dagger$ and $\hat{\gamma}_{2i}=i(\hat{c}_i-\hat{c}_i^\dagger)$. Through these operators we can construct the set of $2^{2N}$ \emph{Majorana strings}:
\begin{equation*}
    \hat{M}_\textbf{v} = i^{\textbf{v}^T\omega_L \textbf{v}}\hat{\gamma}_1^{v_1}\hat{\gamma}_2^{v_2} ... \hat{\gamma}_{2N-1}^{v_{2N-1}}\hat{\gamma}_{2N}^{v_{2N}},
\end{equation*}
where \textbf{v}$\in (\mathbb{Z}_2)^{2N}$ is a binary vector of length $2N$ whose components indicate whether the correspondent Majorana fermion is present or not, while the phase factor $i^{\textbf{v}^T\omega_L \textbf{v}}$ is needed to ensure hermitianicity. The matrix $\omega_L$ has elements equal to one in the lower triangle and zero everywhere else. 
It was shown in \cite{MajoranaCliffordGroup} that Majorana strings are in one-to-one correspondence with Pauli strings, which we denote $P_j$ ($1\leq j\leq4^N$) and can be obtained as generic tensor products of local Pauli operators, namely $P_j \in \{ \mathbb{\hat{1}}_i, \hat{X}_i, \hat{Y}_i, \hat{Z}_i \}^{\otimes N}$. For this reason, they form a complete orthogonal basis for the space of operators, since $\text{Tr}\left[\hat{M}_\textbf{v}\hat{M}_{\textbf{v}\prime}\right]=d\,\delta_{\textbf{v}\textbf{v}\prime}$. Furthermore, we can define the Majorana-Clifford group $\mathcal{C}_N$ as the group of unitary operators that map Majorana strings to Majorana strings, as follows
\begin{equation*}
    \mathcal{C}_N=\{U: U^\dagger \hat{M}_\textbf{v} U = \hat{M}_{\textbf{v}\prime}\}.
\end{equation*}

\begin{figure}[htbp]
  \centering
  \includegraphics[width=\linewidth]{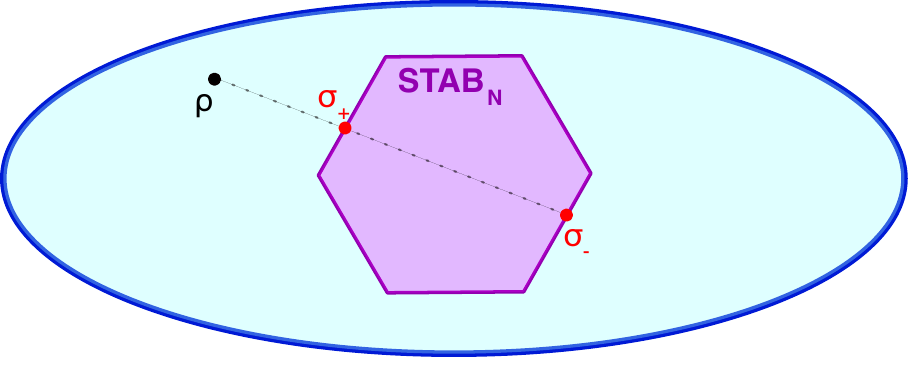}
  \caption{Geometric interpretation of the robustness of magic. The hexagon represents the stabilizer polytope STAB$_N$ for $N=1$, i.e. all possible convex combinations of single qubit stabilizer states. Its vertices correspond to pure stabilizer states. The highlighted points correspond to the two (mixed) stabilizer states $\sigma_\pm$ such that the decomposition $\rho = (1+s)\sigma_+ -s\sigma_-$ is optimal. This is analogous to the formulation given in \cref{eq: RoM}.}
  \label{fig: StabilizerPolytope}
\end{figure}

The states that can be obtained by means of Clifford operations starting from the vacuum state $|0\rangle^{\otimes N}$ are called \emph{stabilizer states} \cite{gottesmanstabilizercodes}. We denote as $\mathcal{S}_N$ the set of all pure N-qubit stabilizer states, whose number of elements is \cite{stabilizercircuits,stabilizergeometry}
\begin{equation*}
    |\mathcal{S}_N| = 2^N\prod^N_{k=1}(2^k+1), 
\end{equation*}
which is super-exponential in the number of qubits. We further define the convex hull of stabilizer states, as the classical mixture of pure stabilizer states
\begin{equation*}
    \text{STAB}_N = \left\{ \ \sum_{i=1}^{|\mathcal{S}_N|} p_i\sigma_i \ \Big| \ \sigma_i \in \mathcal{S}_N, \; p_i \geq 0, \; \sum_i p_i = 1 \ \right\},
\end{equation*}
which constitutes the stabilizer polytope illustrated in \cref{fig: StabilizerPolytope}, and contains all mixed stabilizer states \cite{PolytopeSymmetries,RomMulti-qubit}.

Clifford operations can be generated using only Hadamard, $\pi/2$ phase, and controlled-not gates \cite{stabilizercircuits}. Although circuits composed exclusively of Clifford operations can produce arbitrarily large amounts of entanglement, they can nonetheless be efficiently simulated on classical computers via the Gottesman–Knill theorem \cite{gottesmantheorem}. As a consequence, Clifford circuits alone are not sufficient for universal quantum computation and cannot provide any quantum computational advantage \cite{kitaevuniversalQC}. The same limitation applies to quantum states belonging to the stabilizer polytope STAB$_N$. Indeed, if a state $\rho$ admits a convex decomposition of the form
\begin{equation*}
    \rho = \sum_i p_i \sigma_i
\end{equation*}
then one may efficiently mimic a quantum computer by classically sampling the stabilizer $\sigma_i$ with probability $p_i$ and simulate its subsequent evolution using the Gottesman–Knill protocol \cite{stabilizergeometry,PolytopeSymmetries}. Hence, mixed stabilizer states are equally useless as computational resources.

It therefore becomes essential to quantify how far a given quantum state lies from the stabilizer polytope. In this context, non-stabilizerness—often referred to as magic—emerges as an additional resource required to achieve quantum computational advantage, as it measures the distance between a given state and STAB$_N$. Motivated by this perspective, in the following sections we focus on two measures that have been proposed to quantify magic, with particular emphasis on whether or not they are well defined for mixed quantum states.

\subsection{Robustness of magic (\LR)}
\label{sec: ROM}

The resource theory of magic can be developed in close analogy to the well-established concept of \emph{robustness of entanglement}, that was put forward in Ref. \cite{RobustnessOfEntanglement}. This quantity generally quantifies the endurance of entanglement against noise, by accurately quantifying the minimal mixing with a separable state that is required to completely wash out entanglement. In the context of non-stabilizerness, one can proceed analogously by considering the convex hull of stabilizer states as the set of \emph{free states}. 

Since pure stabilizer states $\sigma_i \in \mathcal{S}_N$ form an overcomplete basis for the set of d-dimensional matrices, we can write any density matrix as an affine combination of pure stabilizer states $\rho = \sum_{i=1}^{|\mathcal{S}_N|} x_i\sigma_i$. In this expression, the coefficients $x_i$ of the decomposition form a quasi-probability distribution, as they satisfy $\sum_i x_i = 1$ but may be negative. Furthermore, the vector $\bm{x}$ is not unique. Then, the \emph
{robustness of magic} $\mathcal{R}$ is defined as the minimal $l_1$-norm $||\bm{x}||_1 = \sum_i|x_i|$ over all possible decompositions \cite{PolytopeSymmetries,manybodyquantummagic,ROMproperties}
\begin{equation}
    \label{eq: RoM}
    \mathcal{R}(\rho) = \min_{\bm{x}} \left\{\, ||\bm{x}||_1 \, \; \Big| \; \rho = \sum_{i=1}^{|\mathcal{S}_N|} x_i\sigma_i, \; \sigma_i \in \mathcal{S}_N \, \right\},
\end{equation}
and quantifies the minimum overlap between the density matrix $\rho$ and the stabilizer polytope. The $l_1$-norm
\begin{equation*}
    ||\bm{x}||_1 = \sum_{i=1}^{|\mathcal{S}_N|}|x_i| = 1 + 2\sum_{i; \, x_i<0}|x_i|
\end{equation*}
measures the amount of negativity in the affine decomposition, which has been related to the simulation runtime in the context of quantum computation. Indeed, in Monte Carlo simulation, the number of samples needed to achieve a particular accuracy scales as $\mathcal{O}\left(\mathcal{R}(\rho)^2\right)$ \cite{QMCNegativity,ROMproperties}.

By collecting terms with the same sign in the above expression, one can express the density matrix as a combination of two (mixed) stabilizer states and a real positive number, as 
\begin{equation*}
    \rho = (1+s)\sigma_+-s\sigma_-,
\end{equation*}
as shown in \cref{fig: StabilizerPolytope}. The robustness of magic is then equivalently defined as 
\begin{equation*}
    \mathcal{R}(\rho) = \min_{\sigma_\pm\in \text{STAB}_N} \Bigg\{ \, 2s+1 \quad \Big| \quad  \rho = (1+s)\sigma_+-s\sigma_-, \, s\geq0 \, \Bigg\},
\end{equation*}
which has a clean geometrical interpretation (see \cref{fig: StabilizerPolytope}) and represents the minimum weight of the combination of stabilizers that can reproduce the state $\rho$.

The problem of \cref{eq: RoM} can be equivalently reformulated as the minimization over the solutions of the system of linear equations
\begin{equation*}
    \mathcal{R}(\rho) = \min_{\bm{x}} \left\{\, ||\bm{x}||_1 \, \; \Big| \; \bm{A}_N \bm{x} = \bm{b} \right\},
\end{equation*}
where we have used the unique decomposition of the quantum state onto the basis of N-qubit Pauli strings, which leads to the definitions $b_j = \text{Tr}(\rho P_j)$ and $(\bm{A}_N)_{ij} = \text{Tr}(\sigma_i P_j)$, where $P_j$ is the j-th Pauli string ($1\leq j\leq4^N$). In this way, one can relate the optimization over the large set of stabilizer states to a standard problem of linear algebra \cite{HandbookRoM}.

The robustness of magic is a good magic measure, in the sense that it satisfies all the required properties. Indeed, it is faithful, meaning that $\mathcal{R}(\rho)\geq 1$, and it is equal to 1 if and only if $\rho$ belongs to the convex hull of stabilizers. As a magic monotone, it is non-increasing under all trace preserving stabilizer channels $\mathcal{E}$, namely $\mathcal{R}(\mathcal{E}(\rho))\leq \mathcal{R}(\rho)$. Furthermore, it is submultiplicative, i.e. $\mathcal{R}(\rho_1 \otimes \rho_2)\leq \mathcal{R}(\rho_1) \mathcal{R}(\rho_2)$ and convex, $\mathcal{R}(\sum_k p_k\rho_k)\leq \sum_k |p_k|\mathcal{R}(\rho_k)$. Together, these properties ensure that $\mathcal{R}$ is well-behaved for arbitrary quantum states.

By taking the logarithm of $\mathcal{R}$ one obtains the \emph{log-free robustness of magic}
\begin{equation}
    \label{eq: LRoM}
    \LR(\rho) = \log_2\left(\mathcal{R}(\rho)\right)
\end{equation}
where we used the base-2 logarithm so that entropies are expressed in units of bits. Inheriting the properties of $\mathcal{R}$, the log-free robustness of magic is also a valid measure in the context of magic resource theory. In particular, it is subadditive, meaning $\LR(\rho_1 \otimes \rho_2)\leq \LR(\rho_1) + \LR(\rho_2)$, and satisfies $\LR(\rho)\geq 0$, where the equality holds only for stabilizer states.

\subsection{Stabilizer Renyi entropy (SRE)}
\label{sec: SRE}

Despite being a valuable measure of non-stabilizerness for mixed states, $\LR(\rho)$ involves computationally hard optimizations over the large set of stabilizers, limiting its applicability to systems of size $N\lesssim 5$ qubits. To overcome this, the 
stabilizer Rényi entropy (SRE) has been introduced \cite{SREalioscia}, and has been proven to be significantly more tractable in large many-body systems, where it can be estimated using sampling-based techniques developed for Majorana and Pauli strings \cite{mariosampling,PoetriGauge,mariofermionicgaussian}.

For a pure quantum state $\rho$, the SRE is defined as the classical Rényi entropy of order $\alpha$ associated with the probability distribution $\pi_\rho(\textbf{v}) = \text{Tr}^2(\rho \hat{M}_\textbf{v}) / d$ \cite{SREalioscia}
\begin{equation}
\label{eq: SRE}
\mathcal{M}_\alpha(\rho) = \frac{1}{1-\alpha} \log_2 \left[ \sum_{\textbf{v} \in (\mathbb{Z}_2)^{2N}} \pi^\alpha_\rho(\textbf{v}) \right] - N,
\end{equation}
up to a constant shift. This entropy captures how broadly the quantum state $\rho$ is distributed over the complete basis of Majorana strings $\hat{M}_\textbf{v}$.

Due to its efficiency and scalability, the SRE has been successfully applied to various many-body systems, including spin chains \cite{aliosciaising,SreRydberg,SreFrustration,SreFazio,SreKivelson,SreViscardi}, fermionic states \cite{mariofermionicgaussian,SYK,magicmolecules}, lattice gauge theories \cite{PoetriGauge,SreGravity}, neural quantum states \cite{SreMelloNeuralNetworks,SreMelloNeuralStates}, matrix product states \cite{SreMpsLami,SreMpsHaug,SreMpsPoetri} and even experimental realizations \cite{AliosciaMeasure,SreMeasure,SreMeasure2,SreMeasure3,TurkeshiMeasure}.
Despite this broad applicability, the SRE is well defined only in the context of pure states.

While for the special case $\alpha = 2$ the SRE has been extended to a subset of mixed states \cite{SREalioscia}:
\begin{equation}
\label{eq: mixedSRE}
\tilde{\mathcal{M}}_2(\rho) \equiv \mathcal{M}_2(\rho) - \mathcal{S}_2(\rho) = -\log_2 \left[ \frac{ \sum_\textbf{v} \text{Tr}^4\left(\rho \hat{M}_\textbf{v}\right) }{ \sum_\textbf{v} \text{Tr}^2\left(\rho \hat{M}_\textbf{v}\right) } \right],
\end{equation}
where $\mathcal{S}_2(\rho) = -\log_2 \text{Tr}(\rho^2)$ is the  2-Rényi entropy of $\rho$, this correction term only partially accounts for the state's mixedness, as we detail below.

As in entanglement theory, where the von Neumann entropy characterizes entanglement only for pure states, the extension of the SRE to mixed states encounters fundamental limitations. Indeed, \cref{eq: mixedSRE} only quantifies deviations from stabilizer states of the form \cite{SREalioscia}
\begin{equation}
\label{eq: stabSRE}
\rho = \frac{1}{d} \left( \mathbb{1} + \sum_{\hat{M}_\textbf{v} \in G} \phi_\textbf{v} \hat{M}_\textbf{v} \right),
\end{equation}
where $\phi_\textbf{v} = {\pm 1}$ and $G$ is a subgroup of the full set of Majorana strings ($|G| < d - 1$). However, \cref{eq: stabSRE} fails to capture the full set of mixed stabilizers, which corresponds to the convex hull of pure stabilizer states. Consequently, \cref{eq: mixedSRE} is a well defined magic measure only for a restricted subset of mixed states, and generally yields an overestimation when applied more broadly. In this sense, it cannot be considered a faithful measure for mixed states—a serious limitation, whose consequences will be explored in the upcoming sections.

\section{Model}
\label{sec: Model}

We consider the two-site Hubbard model at half-filling, defined by the following Hamiltonian
\begin{equation}
\label{eq: HDimer}
    \mathcal{H} = -t\sum_{\sigma\in\{\uparrow,\downarrow\}} \left( \hat{c}^\dagger_{1,\sigma}\hat{c}_{2,\sigma} + \hat{c}^\dagger_{2,\sigma}\hat{c}_{1,\sigma} \right) + U \sum_{i=1,2}\hat{n}_{i,\uparrow}\hat{n}_{i,\downarrow},
\end{equation}
often referred to as the \emph{Hubbard dimer}. The operator $\hat{c}_{i\sigma}$ ($\hat{c}^\dagger_{i\sigma}$) annihilates (creates) an electron with spin $\sigma$ at the site $i$ of the lattice, $\hat{n}_{i\sigma} = \hat{c}^\dagger_{i\sigma} \hat{c}_{i\sigma}$ is the local spin-resolved density, $t$ is the nearest-neighbor hopping amplitude, and $U$ is the local Coulomb repulsion. \\
When extended to an infinite lattice, the Hamiltonian \cref{eq: HDimer} hosts the celebrated Mott metal-to-insulator transition as the ratio $U/t$ increases. Since in our case we only consider two sites, one cannot really speak of a phase transition, but nonetheless we can identify two distinct regimes: for $U\ll t$ electrons will be well delocalized on the whole dimer, while for $U\gg t$ they strongly localize on either of the two  sites. 

Even though an analytical solution for the Hubbard model on an infinite lattice is currently not available -- except for the limits of one and infinite dimensions, for the Hubbard dimer we can derive an exact, analytical expression for the eigenstates restricted to the sector with $N_\uparrow=N_\downarrow=1$, since the Hamiltonian conserves the total number of particles and magnetization. In the basis $| n_{1,\uparrow} n_{1,\downarrow} n_{2,\uparrow} n_{2,\downarrow}  \rangle$ the unique ground state reads
\begin{equation}
\label{eq: gs}
    \left|\psi_-\right\rangle = \frac{1}{\mathcal{N}_+}\left( \left|\uparrow\downarrow,\circ\right\rangle + \Delta_+ \left|\uparrow,\downarrow\right\rangle - \Delta_+ \left|\downarrow,\uparrow\right\rangle + \left|\circ,\uparrow\downarrow\right\rangle \right),
\end{equation}
while the excited states are
\begin{equation}
\label{eq: excited_states}
\begin{aligned}
    \left|\psi_+\right\rangle &= \frac{1}{\mathcal{N}_-}\left( \left|\uparrow\downarrow,\circ\right\rangle + \Delta_- \left|\uparrow,\downarrow\right\rangle - \Delta_- \left|\downarrow,\uparrow\right\rangle + \left|\circ,\uparrow\downarrow\right\rangle \right) \\
    \left|D\right\rangle &= \frac{1}{\sqrt{2}}\left( \left|\uparrow\downarrow, \circ\right\rangle - \left|\circ, \uparrow\downarrow\right\rangle \right) \\
    \left|t_0\right\rangle &= \frac{1}{\sqrt{2}}\left( \left|\uparrow,\downarrow\right\rangle + \left|\downarrow,\uparrow\right\rangle \right)
\end{aligned}
\end{equation}
where we have defined $\Delta_\pm = U / 4t \pm \sqrt{1+(U/4t)^2} $, while $\mathcal{N_\pm}=\sqrt{2(1+\Delta_\pm^2)}$ is a suitable normalization factor. The corresponding eigenenergies are $\{E_-, E_+, U, 0\}$, where we have defined $E_\pm = U/2 \pm 2t\sqrt{1+(U/4t)^2}$. Finally, we point out that all eigenstates can be written in the occupation basis of the local spin-orbitals $| n_{i,\uparrow}\rangle \otimes | n_{i,\downarrow} \rangle$, via the identification $\left|\circ\right\rangle = \left|0\right\rangle\otimes\left|0\right\rangle$, $\left|\uparrow\right\rangle = \left|1\right\rangle\otimes\left|0\right\rangle$, $\left|\downarrow\right\rangle = \left|0\right\rangle\otimes\left|1\right\rangle$, $\left|\uparrow\downarrow\right\rangle = \left|1\right\rangle\otimes\left|1\right\rangle$.

\section{Zero temperature}
\label{sec: ZeroT}

\begin{figure}[htbp]
  \centering
  \includegraphics[width=\linewidth]{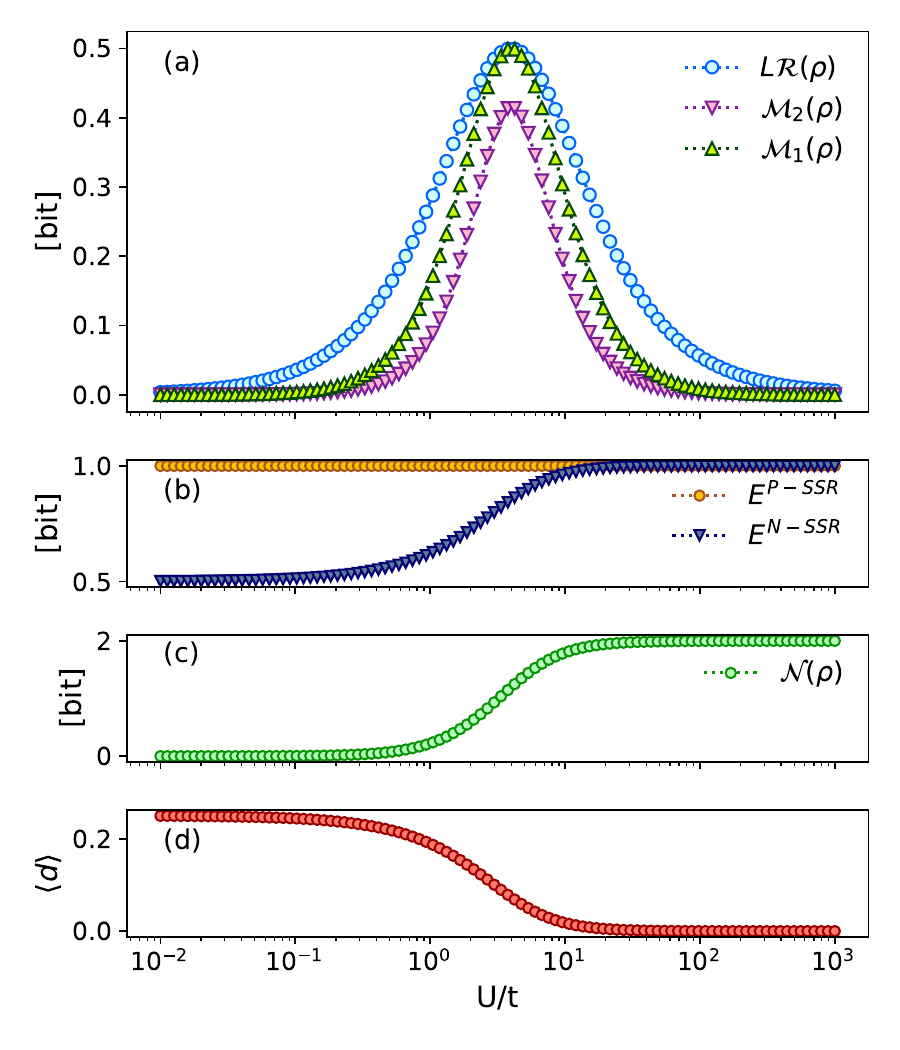}
  \caption{Non-stabilizerness measures (panel a), inter-site entanglement under parity (P-SSR) and charge (N-SSR) super-selection rules (panel b), non-Gaussianity per site (panel c) and double occupancy (panel d) as functions of $U/t$ for the pure ground state of the Hubbard dimer at zero temperature. Non-stabilizerness, entanglement and non-Gaussianity are expressed in bits (units of $\log(2)$).}
  \label{fig: HubbardDimer}
\end{figure}

We start by considering the system at zero temperature. Clearly, in this case all physical properties are solely determined by the pure ground state $|\psi_-\rangle$ of \cref{eq: gs}. For instance, the double occupancy per site, i.e.\,the probability for a site to be occupied simultaneously by two electrons with opposite spin, can be computed as
\begin{equation*}
    \langle d \rangle = \frac{1}{2}\left \langle \psi_-|\hat{n}_{1\uparrow}\hat{n}_{1\downarrow} + \hat{n}_{2\uparrow}\hat{n}_{2\downarrow}|\psi_-\right\rangle= \frac{1}{\mathcal{N}_+^2}.
\end{equation*}
This quantity, which is crucial to asses the mobility of electrons within the dimer, is shown in panel (d) of \cref{fig: HubbardDimer}: as the Coulomb repulsion grows, double occupancies are greatly suppressed, signaling the onset of the aforementioned charge localization on the dimer.

Owing to the fact that the system under study can be considered as a collection of four qubits, we can thoroughly analyze its quantum magic for a wide range of values of $U/t$. We display the $\LR$ and the 2-SRE and 1-SRE computed on the pure ground state $|\psi_-\rangle$ in the top panel of \cref{fig: HubbardDimer}. Remarkably, all quantities exhibit a pronounced peak at intermediate values of $U/t$, and vanish in both the weakly interacting and strongly interacting limits. This behavior aligns with physical intuition: in the non-interacting limit ($U = 0$), the wavefunction is completely delocalized over the dimer, and the system is well described by single-particle states. 
In the opposite limit ($U \gg t$), charge fluctuations are entirely suppressed, and the system reduces to a spin-singlet, which is a stabilizer, despite being maximally entangled. Both extremes do not support any significant degree of quantum complexity. 
It is therefore natural that the non-stabilizerness reaches its maximum at intermediate interaction strengths, where the system deviates most strongly from both limits. We also observe that, while the fully localized state in the limit $U\rightarrow\infty$ does not retain any quantum magic, the maximum of non-stabilizerness occurs in conjunction with a strong depletion of doubly occupied states, indicating that the maximum non-stabilizerness is tightly bound to the onset of localization on the dimer.
To provide a more complete assessment of quantum resources in the model, we compute inter-site entanglement and non-Gaussianity as functions of the Coulomb repulsion $U$ (panels (b) and (c) of \cref{fig: HubbardDimer}, respectively).
While for this two-site pure ground-state one could readily evaluate the inter-site entanglement by computing the local von Neumann entropy \cite{zanardi2002quantum,LocalEntropy_Franca,entropy_MottMIT,walsh2019local}, if one wants to single out the portion of the two-site entanglement that constitutes a genuine quantum resource, appropriate superselection rules must be applied \cite{Banuls_SSR,Friis2013,Friis2016,SchillingParadox,SchillingInterorbital,SchillingReview,SchillingSSRs,bellomia2024quasilocal}.
While for mixed states this minimization has no general analytical solution, in this case, as one starts from a pure state and the reduced states are single orbitals, one obtains closed formulas for the parity and charge superselected entanglement between the sites of the Hubbard dimer \cite{SchillingInterorbital,SchillingParadox}, which we denote $E^\text{P-SSR}$ and $E^\text{N-SSR}$ respectively:
\begin{align}
    E^\text{N-SSR} &= \left(1-2\langle d \rangle\right) \log{2} , \label{eq: N-SSR}\\
    E^\text{P-SSR} &= \log{2}.
    \label{eq: P-SSR}
\end{align}
Analogously, non-Gaussianity, rigorously quantifies how much a state deviates from the set of non-interacting states, and is formally defined as the relative entropy between the given state and the closest Gaussian state. Remarkably, this minimization over the set of Gaussian states can be carried out analytically, so that the non-Gaussianity of an arbitrary state $\rho$ can be expressed as \cite{gottlieb0,gottlieb1,gottlieb2}
\begin{equation}
    \mathcal{N}(\rho) = s(\gamma_\rho) + s(\mathbb{I}-\gamma_\rho) - s(\rho),
    \label{eq: Nonfreeness}
\end{equation}
where $(\gamma_\rho)_{ij} = \langle c_j^\dagger c_i\rangle_\rho$ is the one-body density matrix of the system, and $s(\rho)$ is the von Neumann entropy. Non-Gaussianity quantifies in a well-defined way correlations between electrons \cite{gottlieb0,gottlieb1,gottlieb2,vollhardtnonfreeness,Held_nonfreeness}, constitutes a genuine resource of state-complexity \cite{Paris_nongaussianity,Paris_2,Paris_3} and has been connected to the information theory of many-body systems \cite{schillingnaturalorbitals,Bellomia_intracorr,Zavatti2025}.

Our results highlight the striking difference between these quantities and magic. Both entanglement and non-Gaussianity, indeed, grow monotonically with the Coulomb repulsion, and saturate to a constant value in the limit of large $U$. In particular, the non-Gaussianity per site, displayed in \cref{fig: HubbardDimer}(c) is zero only in the non-interacting case $U=0$, while the entanglement (\cref{fig: HubbardDimer}(b)) is everywhere larger than zero. 
Interestingly, both entanglement and non-Gaussianity become maximal in the large-$U$ limit, where the ground state of the model is a stabilizer. This proves that magic, entanglement and non-Gaussianity probe complementary aspects of complexity.

\subsection{Absence of local magic}
\label{sec: LocalCase}

We now turn our attention to an analysis of local resources in the Hubbard dimer, which provides a first insight on the unfaithfulness of the SRE. Starting from the ground state density matrix $|\psi_-\rangle\langle\psi_-|$ of the Hubbard dimer, we can consider the density matrix of one site (its \emph{local} component) by tracing out the other one, yielding the definition of the \emph{local reduced density matrix} (LRDM):
\begin{equation*}
    \rho_1 = \text{Tr}_2 \left(|\psi_-\rangle\langle\psi_-|\right).
\end{equation*}
The LRDM represents the density matrix of a single fermionic orbital, and can therefore be expressed in the occupation basis of its spin-orbitals $| n_{1,\uparrow}\rangle \otimes | n_{1,\downarrow} \rangle$, as described in \cref{sec: Model}. In this basis, $\rho_1$ takes the form
\begin{align}
    \rho_1 & = \frac{1}{\mathcal{N}_+^2} \left(|\circ\rangle\langle\circ\right|
    +\Delta_+^2\left|\uparrow\rangle\langle\uparrow\right|
    +\Delta_+^2\left|\downarrow\rangle\langle\downarrow\right|
    +\left|\uparrow\downarrow\rangle\left\langle\uparrow\downarrow\right|\right) \nonumber \\ 
    & = \langle d \rangle \left(\left|\circ\rangle\langle\circ\right|
    + \left|\uparrow\downarrow\rangle\langle\uparrow\downarrow\right| \right) +  \nonumber \\
    & \qquad \qquad \qquad \qquad + \left(\frac{1}{2}-\langle d \rangle \right)
    \left(\left|\uparrow\rangle\langle\uparrow\right|+\left|\downarrow\rangle\langle\downarrow\right|\right), \nonumber \\
    & = \langle d \rangle \left(\left|0,0\rangle\langle0,0\right|
    + \left|1,1\rangle\langle1,1\right| \right) +  \nonumber \\
    & \qquad \quad + \left(\frac{1}{2}-\langle d \rangle \right)
    \left(\left|1,0\rangle\langle1,0\right|+\left|0,1\rangle\langle0,1\right|\right),
    \label{eq: localRDM}
\end{align}
where in the second line we have recasted the expression in terms of the double occupancy, using the fact that $\langle d\rangle = 1/\mathcal{N}^2_+$ and $1/2-\langle d\rangle = \Delta^2_+/\mathcal{N}^2_+$, and in the third line we have explicitly written it in the spin-orbital occupation basis. 

\begin{figure}[htbp]
  \centering
  \includegraphics[width=\linewidth]{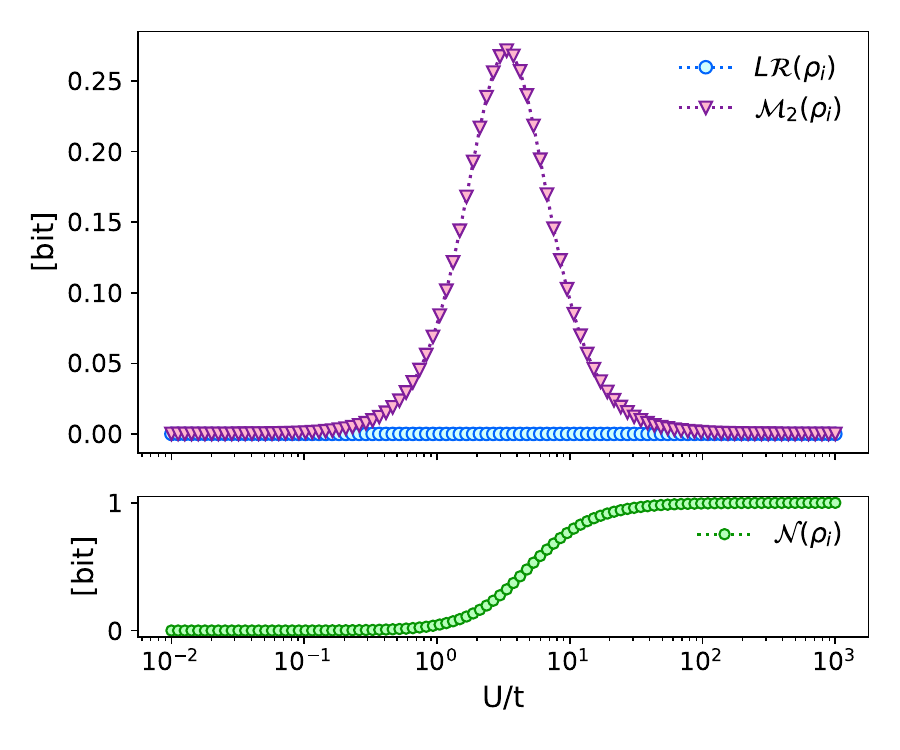}
  \caption{Top panel: Local non-stabilizerness as a function of $U/t$, measured for the one-site reduced density matrix of the Hubbard dimer. While the system is clearly devoid of local magic, this is correctly captured only by $\LR$, while the 2-SRE is everywhere non-zero. Bottom panel: local non-Gaussianity. Unlike magic and entanglement, non-Gaussianity is zero only at $U=0$, highlighting the different nature of these quantities.}
  \label{fig: HubbardDimerLocal}
\end{figure}

The form of \cref{eq: localRDM} explicitly shows that the LRDM is diagonal and uniquely composed of two-qubit stabilizers, since all states that appear in $\rho_1$ can be obtained as tensor products of $\left|0\rangle\langle0\right|$ and $\left|1\rangle\langle1\right|$, which are single-qubit stabilizer states.
It is therefore clear that for each value of $U/t$ we are in the presence of a classical mixture of stabilizer states, which does not host any quantum magic. 
The log-free robustness of magic captures this by vanishing exactly, unambiguously marking the absence of local quantum magic, as shown in \cref{fig: HubbardDimerLocal}(a).

We stress here that the diagonal form of the LRDM is a completely general feature of Hubbard-like Hamiltonians with U$(1)$ and SU$(N)$ symmetries ($N$ being the number of fermionic flavours, in the most popular single-orbital Hubbard model $N=2$), even on an infinite lattice, and is valid also away from half filling with minor modifications \cite{zanardi2002quantum,Bellomia_intracorr,Zavatti2025}.
Hence, this result generally proves the absence of local magic for this family of models,
suggesting that \emph{nonlocal magic} \cite{gravitationalbackreaction,nonlocalmagic}
characterizes these physical systems. In formal terms, nonlocal magic quantifies the amount of non-stabilizerness which cannot be erased with purely local operations acting separately on two subsystems. For the special case of the two-site Hubbard model, this results in 
\begin{equation}
    \mathcal{M}^\mathrm{NL}(|\psi\rangle) = \min_{U_1,U_2}\mathcal{M}(U_1\otimes U_2 |\psi\rangle),
\label{eq: nonlocal magic}
\end{equation}
where $\mathcal{M}$ is any faithful magic monotone and $U_1$ and $U_2$ are unitaries acting on sites 1 and 2 respectively. Our analysis on the absence of non-stabilizerness in the LDRM suggests that, for the Hubbard dimer, $\mathcal{M}^\mathrm{NL}(|\psi\rangle) = \mathcal{M}(|\psi\rangle)$, namely that the non-stabilizer content is purely nonlocal. In Appendix \ref{app: nonlocal magic}, by performing the minimization underlying \cref{eq: nonlocal magic}, with $\mathcal{M}$ given as $\mathcal{M}_2$, we explicitly show that this is indeed the case, rigorously proving the absence of local non-stabilizerness also in the SRE framework.

The vanishing of local magic aligns with the fact that entanglement is also absent at the local level. Indeed, \cref{eq: localRDM} represents a mixture of separable states, therefore ruling out the possibility of intra-orbital entanglement. It follows that all correlations between spin-orbitals are due to the classical mixture of states, and are entirely captured by the local non-Gaussianity \cite{Bellomia_intracorr,Zavatti2025}, which grows monotonically with the Coulomb repulsion and saturates to $\log(2)$ as shown in the bottom panel of \cref{fig: HubbardDimerLocal}. This is the maximum value for classical correlations in a 2-qubit system, in stark contrast with the $2\log(2)$ value of the nonlocal non-Gaussianity per site in \cref{fig: HubbardDimer}, which is instead characteristic of maximal \emph{quantum} non-Gaussian correlations. Once again, a clear signature of the classical nature of local Hubbard states. 

On the other hand, the mixed SRE $\mathcal{\tilde{M}}_2$ behaves very differently from the $\LR$. Indeed, it is everywhere greater than zero and reaches its maximum at intermediate values of $U/t$, as shown in \cref{fig: HubbardDimerLocal}(a). The failure of $\mathcal{\tilde{M}}_2$ in capturing the stabilizer nature of $\rho_1$ arises from the fact that the SRE vanishes only if the density matrix can be written in the form $\rho_1=\tfrac{1}{4}(\hat{\mathbb{1}}+\sum_\textbf{v}\phi_\textbf{v}M_\textbf{v})$ with $\phi_\textbf{v}=\pm 1$ (cf.\,\cref{eq: stabSRE}), which however is not the most general mixed stabilizer state. Indeed, for any value of $U$ the local reduced density matrix takes the form
\begin{equation*}
\rho_1 = \frac{1}{4} \left( \hat{\mathbb{1}} - \delta \hat{P} \right),
\end{equation*}
where $\delta = \frac{2(\Delta^2_+ - 1)}{\mathcal{N}^2_+}=1-4\langle d\rangle$ is a real number that depends on $U/t$, not necessarily $\pm 1$, and $\hat{P}$ is the parity operator, i.e. the Majorana string corresponding to the vector $\textbf{v}=(1,1,1,1)$. This leads to
\begin{equation*}
\tilde{\mathcal{M}}_2(\rho_1) = -\log_2 \left( \frac{1+\delta^4}{1+\delta^2} \right),
\end{equation*}
which vanishes only for $\delta=0$ (at $U=0$) and $\delta=1$ (as $U \to \infty$). Due to this clear limitation of the 2-SRE in discriminating between genuine mixed non-stabilizers and states that are convex mixtures of stabilizers, we refrain from computing it in the next section, where we focus on mixed thermal states. 

\section{Finite temperature}
\label{sec: FiniteT}

\begin{figure*}[htbp]
  \centering
  \includegraphics[width=\textwidth]{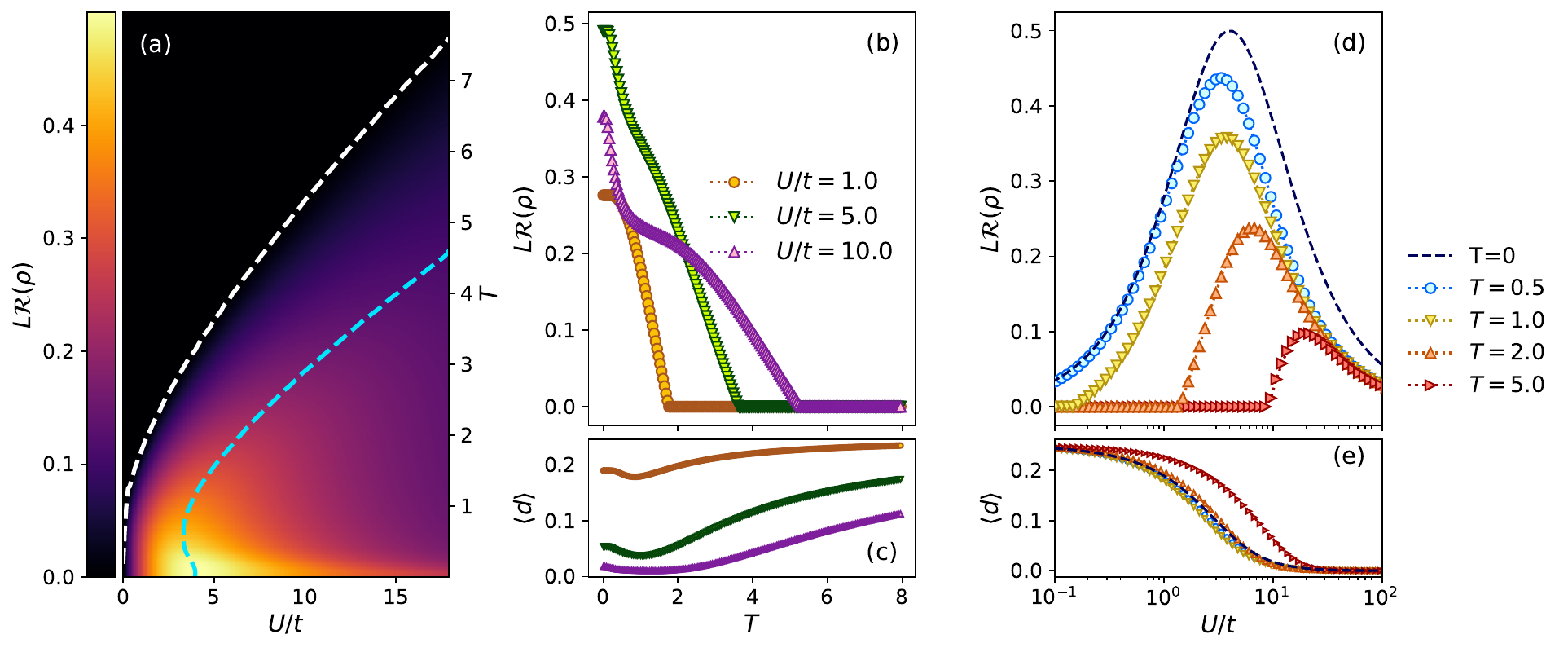}
  \caption{Log-free robustness of magic for the Hubbard dimer at finite temperature. In (a) we display the $U$ vs $T$ phase diagram, showing that non-stabilizerness disappears in the region of large temperature and small Coulomb repulsion. The dashed white line represents the line of critical temperature $T_\mathrm{c}(U/t)$ where the magic drops to zero, while the dashed azure line marks the maximum of non-stabilizerness at each temperature. We further display the evolution of $\LR(\rho)$ and of the double occupancy $\langle d \rangle$ as functions of temperature for fixed Coulomb repulsion in (b) and (c), while in (d) and (e) as functions of $U/t$ in log-scale.}
  \label{fig: Phase_diagram}
\end{figure*}

We now turn our attention to the case of a finite temperature, for which the system is described by the density matrix
\begin{equation*}
    \rho = \frac{e^{-\mathcal{H}/T}}{Z}
\end{equation*}
where $T$ is the temperature in units of $k_B$ and $Z=\sum_i e^{-E_i/T}$ is the partition function. Expanding on the basis of energy eigenstates, the density matrix for fixed $U$ and $T$ reads
\begin{equation}
\label{eq:rho_T}
\begin{aligned}
\rho(U,T) = \frac{1}{Z}\Big(&
|\psi_-\rangle\langle\psi_-|
+ e^{-\frac{E_+ - E_-}{T}} |\psi_+\rangle\langle\psi_+| \\
&+ e^{-\frac{U - E_-}{T}} |D\rangle\langle D|
+ e^{\frac{E_-}{T}} |t_0\rangle\langle t_0|
\Big),
\end{aligned}
\end{equation}
where we used the eigenstates and eigenvalues defined in \cref{eq: gs,eq: excited_states}.

In this situation, the $\LR$ reveals a richer structure in the joint dependence on temperature and Coulomb repulsion, giving rise to a nontrivial phase diagram and to a \emph{thermal death of magic}, similar to the thermal death of entanglement previously pointed out in this kind 
of systems \cite{thermal_concurrence_hubbard_dimer,thermal_spin/charge_concurrence_hubbard_dimer,Canio_concurrence_hubbard_dimer}.
For a fixed ratio $U/t$, increasing the temperature induces a thermal mixing between the non-stabilizer states $|\psi_-\rangle$ and $|\psi_+\rangle$ and the stabilizer states $|D\rangle$ and $|0\rangle$. At sufficiently high temperatures, this mixing drives the thermal state inside the stabilizer polytope, causing the $\LR$ to vanish.
At first sight, one might expect this vanishing to be caused by mixing with the stabilizer states. However, this is not the case: admixture with stabilizer states can only bring the non-stabilizer ground state arbitrarily close to the stabilizer polytope, but never inside it. Instead, the crucial mechanism is the thermal mixing between the two non-stabilizer states $|\psi_-\rangle$ and $|\psi_+\rangle$. As we discuss more thoroughly in Appendix \ref{app:thermal_magic}, when the thermal weight of $|\psi_+\rangle$ becomes sufficiently large, its mixture with the ground state $|\psi_-\rangle$ leads to a (mixed) stabilizer state, thereby eliminating magic.
As a consequence, the $\LR$ vanishes above a critical temperature $T_\mathrm{c}(U/t)$, which depends on the Coulomb repulsion, as highlighted in \cref{fig: Phase_diagram}(b). Moreover, \cref{fig: Phase_diagram}(d) shows that at finite temperature magic is absent for small values of $U/t$ and emerges only when the Coulomb repulsion exceeds a critical value $U_\mathrm{c}$, which can be determined by inverting the relation $T_\mathrm{c}(U)$. Physically, this behaviour reflects the fact that, at small vales of $U$, the contribution of the excited non-stabilizer $|\psi_+\rangle$—which drives the thermal state inside the stabilizer polytope—remains significant, suppressing non-stabilizerness. Only for sufficiently large values of the Coulomb repulsion is the Boltzmann weight of $|\psi_+\rangle$ effectively reduced, allowing magic to develop.

To illustrate how double occupancies shape the $U$ vs $T$ phase diagram, we display in \cref{fig: Phase_diagram}(c) and (e) their evolution as functions of temperature and Coulomb repulsion. The behavior of the double occupancy at high temperatures is straightforward to understand: as $T$ increases, doubly occupied configurations become more thermally populated, causing $\langle d \rangle$ to increase and eventually approach its noninteracting value of 0.25. As a consequence, as shown in \cref{fig: Phase_diagram}(e), double occupancies are suppressed only at low temperatures, while they are generally enhanced at higher temperatures.

At intermediate temperatures, however, the system exhibits a tendency toward localization as the temperature is raised, as shown in \cref{fig: Phase_diagram}(c). The thermal non-monotonicity of the double occupancy has been studied extensively in the half-filled Hubbard model and can be traced back to the fact that localization leads to higher entropy, and is thus thermodynamically favored at intermediate temperatures, until the thermal population of doubly occupied states becomes dominant at higher temperatures \cite{GeorgesInfinitedim,SchaferMultimessenger}.
This non-monotonic behavior underlies the temperature-dependent shift of the maximum of non-stabilizerness, indicated by the dashed azure line in the phase diagram and visible in \cref{fig: Phase_diagram}(d). Indeed, the zero-temperature analysis of \cref{sec: ZeroT} shows that the maximum robustness of magic coincides with strong localization on the dimer. For $T\lesssim 1$, localization on the dimer is thermodynamically favored and sets in already at lower interaction strengths. As a consequence, the position of the maximum in the $\LR$ shifts towards smaller values of $U$ as the temperature is increased. Conversely, at higher temperatures, a strong depletion of double occupancies—and hence the development of large non-stabilizerness—requires progressively larger values of the Coulomb repulsion in order to overcome thermal fluctuations.

\section{Dynamics after an \\ interaction quench}
\label{sec: Dynamics}

\begin{figure*}[htbp]
  \centering
  \includegraphics[width=\textwidth]{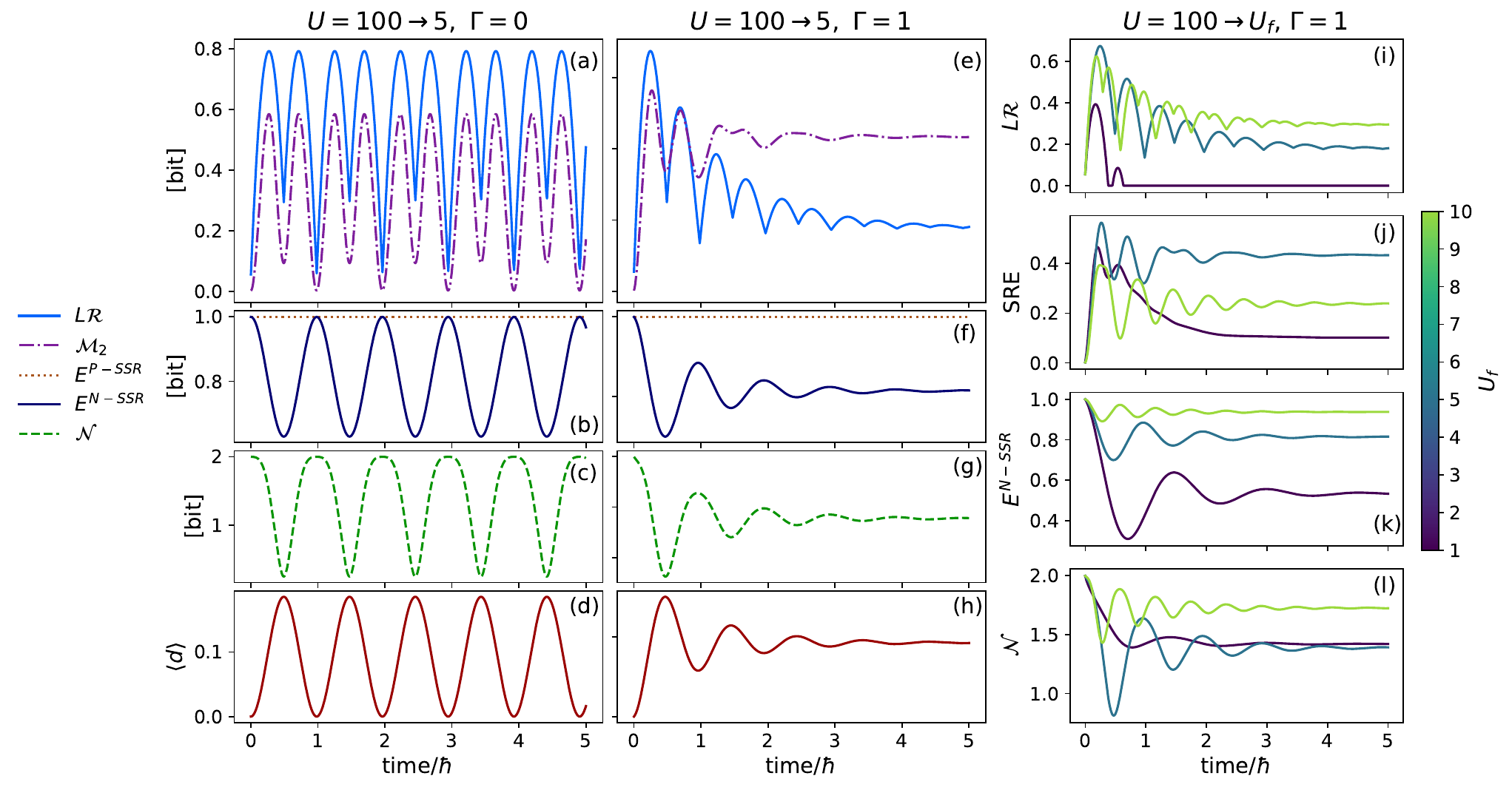}
  \caption{Dynamics of non-stabilizerness after a quench at $t=0$. Panels (a) and (e) display the evolution of $\LR$ and the 2-SRE, (b) and (f) of the inter-site entanglement under SSRs, (c) and (g) of the non-Gaussianity, and finally (d) and (h) of the double occupancy, after a quench from $U=100$ to $U=5$. When dephasing is absent ($\Gamma=0$), non-stabilizerness oscillates periodically, while in presence of dephasing ($\Gamma>0$) it saturates to a constant value.  Panels (i) and (j) show the evolution of $\LR$ and of the 2-SRE respectively, for different values of the final value of the quench $U_\mathrm{f}$. While the 2-SRE always saturates to non-zero values, $\LR$ can become zero after some time, due to the mixing between $|\psi_-\rangle$ and $|\psi_+\rangle$. Finally, panels (k) and (l) show the evolution of the charge super-selected entanglement and non-Gaussianity, respectively: unlike magic, neither saturates to zero at long times.}
  \label{fig: Dynamics_panels}
\end{figure*}

Finally, we study how the quantum dynamics of the Hubbard dimer affect its non-stabilizer content. To this end, we consider a sudden quench in the Coulomb repulsion $U$, from an initial value $U=U_\mathrm{i}$ at times $t<0$ to a final value $U=U_\mathrm{f}$ at times $t\geq0$. Owing to an increasing level of control in cold-atom platforms—especially in optically lattices, where the ratio $U/t$ can be accurately tuned via the laser intensity—this protocol represents a situation of significant physical relevance.
For $t<0$ the Hamiltonian is $\mathcal{H}_0 = \mathcal{H}(U_\mathrm{i})$ and the system is in the ground state $|\psi_0\rangle = |\psi_-(U_\mathrm{i})\rangle$. The quench $U_\mathrm{i} \rightarrow U_\mathrm{f}$ takes place at time $t=0$, and for subsequent times the system evolves as
\begin{equation*}
    |\psi(t)\rangle = e^{-\frac{i}{\hbar} \mathcal{H} \, t}|\psi_0\rangle,
\end{equation*}
under the new Hamiltonian $\mathcal{H} = \mathcal{H}(U_\mathrm{f})$.
To compute the time evolution for $t\geq0$ we rewrite the state $|\psi_0\rangle$ on the basis of eigenstates of $\mathcal{H}$. Since the states $|D\rangle$ and $|t_0\rangle$ do not depend on $U$, they are simultaneously eigenstates of $\mathcal{H}(U_\mathrm{f})$ and $\mathcal{H}(U_\mathrm{i})$, and hence they are always orthogonal to $|\psi_0\rangle$. Then, one is left with

\begin{equation*}
    |\psi_0\rangle = \alpha |\psi_-(U_\mathrm{f})\rangle + \beta |\psi_+(U_\mathrm{f})\rangle,
\end{equation*}
with
\begin{align*}
    \alpha &= \langle \psi_-(U_\mathrm{f})|\psi_0\rangle = \frac{2}{\mathcal{N}_+^0} \left( \frac{1+\Delta^0_+\Delta_+}{\mathcal{N}_+} \right), \\
    \beta &= \langle \psi_+(U_\mathrm{f})|\psi_0\rangle = \frac{2}{\mathcal{N}_+^0} \left( \frac{1+\Delta^0_+\Delta_-}{\mathcal{N}_-} \right),
\end{align*}
where we introduced the following shorthands: $\mathcal{N}_\pm = \mathcal{N}_\pm(U_\mathrm{f})$, $\mathcal{N}^0_\pm = \mathcal{N}_\pm(U_\mathrm{i})$, $\Delta_\pm = \Delta_\pm(U_\mathrm{f})$ and $\Delta^0_\pm = \Delta_\pm(U_\mathrm{i})$.
Finally, the wavefunction evolved at time $t$ reads
\begin{equation}
    |\psi(t)\rangle = \alpha \, e^{-\frac{i}{\hbar} E_- \, t}|\psi_-\rangle + \beta \, e^{-\frac{i}{\hbar} E_+ \, t}|\psi_+\rangle.
\end{equation}

Since the system is closed, the evolution is unitary and the time dynamics exhibit coherent oscillations, which are clearly visible in \cref{fig: Dynamics_panels}(a-d). Interestingly, the quench induces periodic oscillations of the double occupancies, with frequency $(E_+-E_-)/\hbar$. These oscillations, in turn, drive corresponding oscillations in the non-Gaussianity, in the N-SSR inter-site entanglement, as well as in the non-stabilizerness. Moreover, the resulting magic oscillations are twofold in the sense that within each period of the double-occupancy oscillations there are two oscillations of the non-stabilizer content, due the fact that the magic increases when $\langle d \rangle$ deviates from an intermediate value, as shown in \cref{fig: Dynamics_panels}(a).

In real experiments, however, interactions with the environment cause the system to progressively lose coherence, and the oscillatory behaviour discussed above is expected to be damped as the dephasing sets in. To account for this effect, we model the dissipative evolution in the density matrix formalism and apply the following trace-preserving map
\begin{align*}
    |\psi_-\rangle\langle\psi_+| &\rightarrow e^{-\Gamma t} |\psi_-\rangle\langle\psi_+| \\
    |\psi_+\rangle\langle\psi_-| &\rightarrow e^{-\Gamma t} |\psi_+\rangle\langle\psi_-|,
\end{align*}
which induces decoherence during the time evolution when $\Gamma>0$. We observe that $\Gamma^{-1}$ is the corresponding decoherence time.

For any finite value of $\Gamma$, the amplitude of the oscillations is progressively reduced, as shown in \cref{fig: Dynamics_panels}(e-h), and for $t \gg \Gamma^{-1}$ the non-stabilizer content eventually reaches a saturation value.

A primary effect of decoherence is that the system is described by a mixed density matrix $\rho(t)$, which at long times approaches
\begin{equation*}
    \rho(t \gg \Gamma^{-1}) \simeq|\alpha|^2|\psi_-\rangle\langle\psi_-| + |\beta|^2|\psi_+\rangle\langle\psi_+|,
\end{equation*}
corresponding to a time-independent mixture of two non-stabilizer states. As a result, all observables—including measures of non-stabilizerness—saturate to constant values. Moreover, the mixedness of the state implies that the 2-SRE is no longer a faithful measure of magic, as highlighted by the qualitative difference with the $\LR$, shown in \cref{fig: Dynamics_panels}(i) and (j). Interestingly, while the saturation value of the 2-SRE is always positive, this is not generally the case for the $\LR$, which can vanish for specific values of the quench parameter $U_\mathrm{f}$. As we detail in Appendix \ref{app:thermal_magic}, this behavior originates from the fact that particular mixtures of $|\psi_-\rangle$ and $|\psi_+\rangle$ lie inside the stabilizer polytope, and therefore correspond to stabilizer states. Moreover, it is important to notice that unlike magic, entanglement and non-Gaussianity never saturate to zero in presence of dephasing, once again underscoring the difference between these markers of state complexity.
We note in passing that the formulas for the inter-site entanglement under charge and parity SSRs given in \cref{eq: N-SSR,eq: P-SSR} are only valid when the system is described by a pure state, and thus were not used for the case $\Gamma>0$. The correct formulas in the presence of mixed states have been derived in \cite{SchillingSSRs}, and are not reported here for brevity.

\begin{figure}[htbp]
  \centering
  \includegraphics[width=\linewidth]{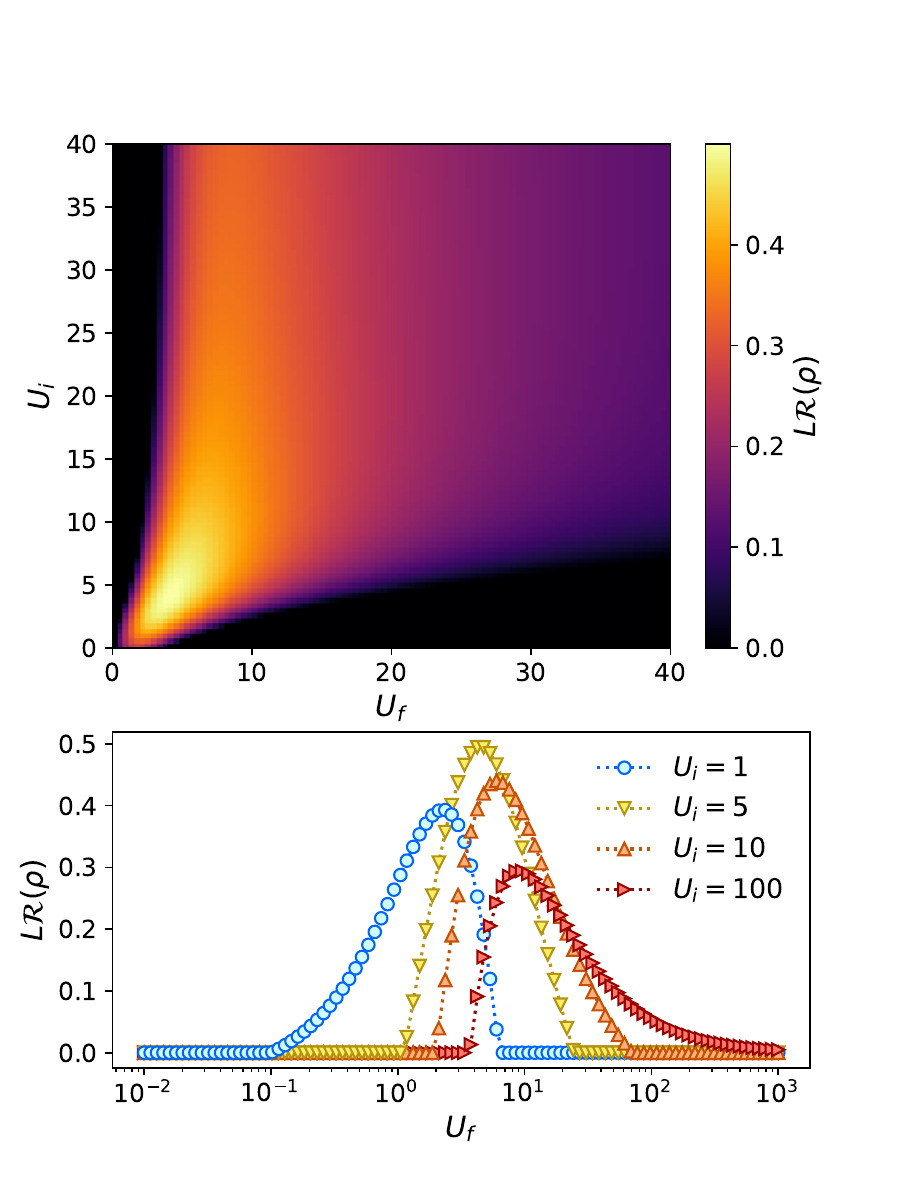}
  \caption{Saturation value of $\LR$ of magic after a quench $U_\mathrm{i} \rightarrow U_\mathrm{f}$ and a transient time $\Gamma t=10$. Top panel: saturation value of  as function of $U_\mathrm{i}$ and $U_\mathrm{f}$. Bottom panel: saturation value for fixed $U_\mathrm{i}$ vs the final values of the quench $U_\mathrm{f}$, corresponding to horizontal cuts in the phase diagram.}
  \label{fig: Dynamics_PD}
\end{figure}

In addition, the saturation value of the $\LR$ as a function of the quench parameters $U_\mathrm{i}$ and $U_\mathrm{f}$, corresponding to the initial and final values of the Coulomb repulsion. This quantity is also experimentally relevant, as it can be interpreted as the long-time average of the non-stabilizer content, 
\begin{equation*}
    \overline{\LR} = \lim_{T\rightarrow\infty} \frac{1}{T}\int_0^T \LR(t) dt .
\end{equation*}
The resulting phase diagram is shown in \cref{fig: Dynamics_PD}. Interestingly, the magic vanishes over a wide range of parameters, indicating that, for these values of $U_\mathrm{i}$ and $U_\mathrm{f}$, the quantum evolution, in the presence of decoherence, drives the system to a (mixed) stabilizer state.

\section{Conclusions and outlooks}
\label{sec: Conclusions}

In this work we have studied the non-stabilizerness (magic) of strongly correlated fermions, using the Hubbard dimer as a paradigmatic model for interacting electrons. This simple, yet non trivial model is solved exactly, providing us with a full access to different measures of quantum magic and other quantum resources in a variety of scenarios, including zero and finite temperature, as well as the non-equilibrium dynamics after a quench of the interaction. Thermal and non-equilibrium protocols call for a proper treatment of mixed states. 

In particular, we have quantified non-stabilizerness using the robustness of magic $\LR$ and the Stabilizer Renyi Entropy SRE. Our results prove that the latter is not a reliable estimate of non-stabilizerness for mixed states. 

Relying on symmetries of the Hubbard Hamiltonian, we proved the completely general result that local magic is absent in these systems, since the local reduced density matrix of the model is always a classical mixture of stabilizer states 
\cite{Bellomia_intracorr}. This property is missed by the SRE, that severely overestimates magic in the local reduced density matrix, for any eigenstate of the model.

By means of $\LR$, we have shown that the nonlocal magic on the dimer grows to a maximal value in correspondence to the onset of electron localization on the dimer at zero temperature, while it becomes zero in the non-interacting and strong-interacting limits. 

A finite temperature analysis, and the dynamics after a quantum quench in the presence of dephasing allow us to solidify and enrich the physical picture. In both cases, we have shown that non-stabilizerness can vanish whenever the mixture of the non-stabilizer energy eigenstates lies within the stabilizer polytope. In the former case, this leads to the thermal death of magic at large temperatures and small Coulomb repulsion while, in the latter, the dynamics can lead the state into the stabilizer polytope at large times, depending on the starting and ending strength of the interaction. Also in this case, the main features are completely missed by the SRE, as they are inherent to the mixed character of the state.

In order to fully assess the role of quantum magic as a marker for complexity in the Hubbard dimer, we have compared it with more traditional tools of quantum information theory, particularly entanglement and non-Gaussianity, which have been intensively studied in the context of strongly correlated fermions and quantum chemistry in the recent years \cite{bellomia2024quasilocal,BippusTwoSiteEntanglement,Held_QFI_pseudodap,SchillingSSRs,SchillingInterorbital,SchillingReview,schillingnaturalorbitals,bellomia2024quantum,Bellomia_intracorr,Held_nonfreeness,vollhardtnonfreeness,Zavatti2025}. Our analysis clearly highlights that non-stabilizerness is a distinct and complementary measure of quantum complexity, necessary in order to provide a complete assessment of available quantum resources in systems of correlated fermions.

Our statement on the absence of local magic holds for the case of the single-orbital Hubbard model and its SU($N$)-symmetric version with $N$ components. Multicomponent models for solids include instead richer interactions describing, e.g., the Hund's exchange, which might lead to the development of non-stabilizerness even at the local level.  Moreover, within nonlocal extensions of dynamical mean-field theory \cite{dmft_review}, it is possible to characterize the quantum complexity of finite clusters \cite{bellomia2024quasilocal,BippusTwoSiteEntanglement}, enabling the characterization of magic for subsystems embedded in the lattice. 

Finally, recent analyses have established the notion of \emph{nonlocal magic}
\cite{gravitationalbackreaction,nonlocalmagic}, 
namely the amount of non-stabilizerness that cannot be removed by a change of the local basis, as the central character in understanding the interplay between magic and entanglement. While remarkable advances have been made towards the understanding of nonlocal magic in Gaussian fermionic systems \cite{nonlocalmagic_gaussian1,nonlocalmagic_gaussian2}, the same cannot be said about interacting fermionic systems. In this framework, our results show that magic in the two-site Hubbard model is decidedly nonlocal, and extending this analysis to larger systems would be important to assess whether this behavior persists in more generic interacting fermionic models.
All in all, our work paves the way to a more complete understanding of quantum resources in the ground and thermal states of fermionic non-Gaussian systems.

\begin{acknowledgements}
We thank M. Collura and G. Lami for insightful discussions.
We acknowledge financial support from the MUR via National Recovery
and Resilience Plan PNRR Projects No.\,CN00000013-ICSC and No.\,PE0000023-NQSTI, as well as via PRIN
2020 (Prot.\,2020JLZ52N-002) and PRIN 2022 (Prot.\,20228YCYY7) programmes. GB further acknowledges
support through the SFB Q-M\&S project of the FWF, DOI 10.55776/F86.
\end{acknowledgements}


\appendix

\section{Nonlocal magic}
\label{app: nonlocal magic}

We devote this appendix to the rigorous study of nonlocal magic, as defined in \cite{gravitationalbackreaction,nonlocalmagic}.
This quantity formally addresses the question of how much of magic is nonlocal, by isolating the amount of non-stabilizerness which cannot be removed via local basis changes. It follows that nonlocal magic is operationally defined as 
\begin{equation}
    \mathcal{M}^\mathrm{NL}\left(|\psi\rangle \right) = \min_{U_1,U_2}\mathcal{M}\left( \, U_1\otimes U_2 |\psi\rangle \, \right),
\label{eq: nonlocal magic app}
\end{equation}
where $\mathcal{M}$ is any faithful measure of magic and $U_1$, $U_2$ are local unitaries that act on sites 1 and 2 respectively. The minimization involved in \cref{eq: nonlocal magic app} is generally very hard, to the point that
even for Gaussian states specially tailored formulas and upper bounds are needed for the quantification of nonlocal magic
\cite{nonlocalmagic_gaussian1,nonlocalmagic_gaussian2}.  

\begin{figure}[h!]
  \centering
  \includegraphics[width=\linewidth]{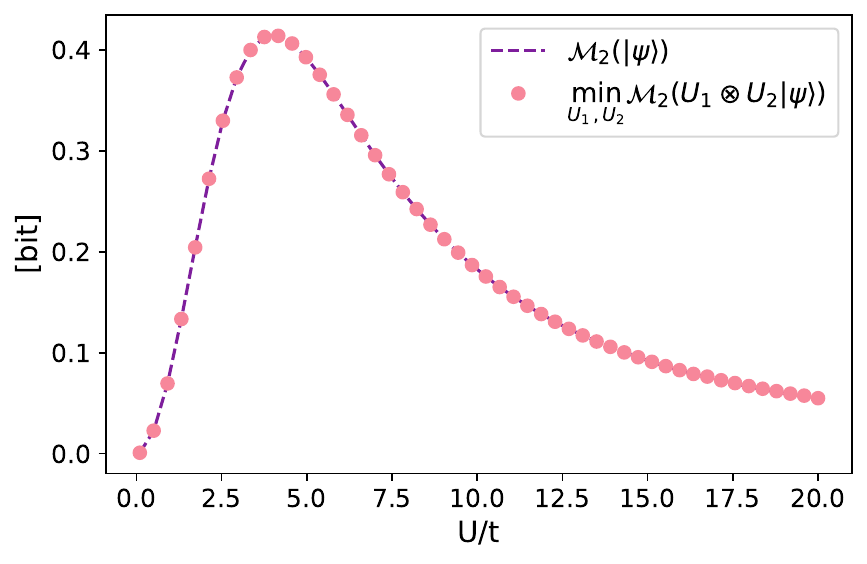}
  \caption{Nonlocal magic (dots) of the ground state of the two-site Hubbard model, measured with the stabilizer Renyi entropy. The comparison with the non-stabilizernees of the ground state (dashed) shows that the two quantities are equivalent, implying the absence of local non-stabilizerness.
  }
  \label{fig: AppendixA}
\end{figure}
\begin{figure*}[ht!]
  \centering
  \includegraphics[width=.95\textwidth]{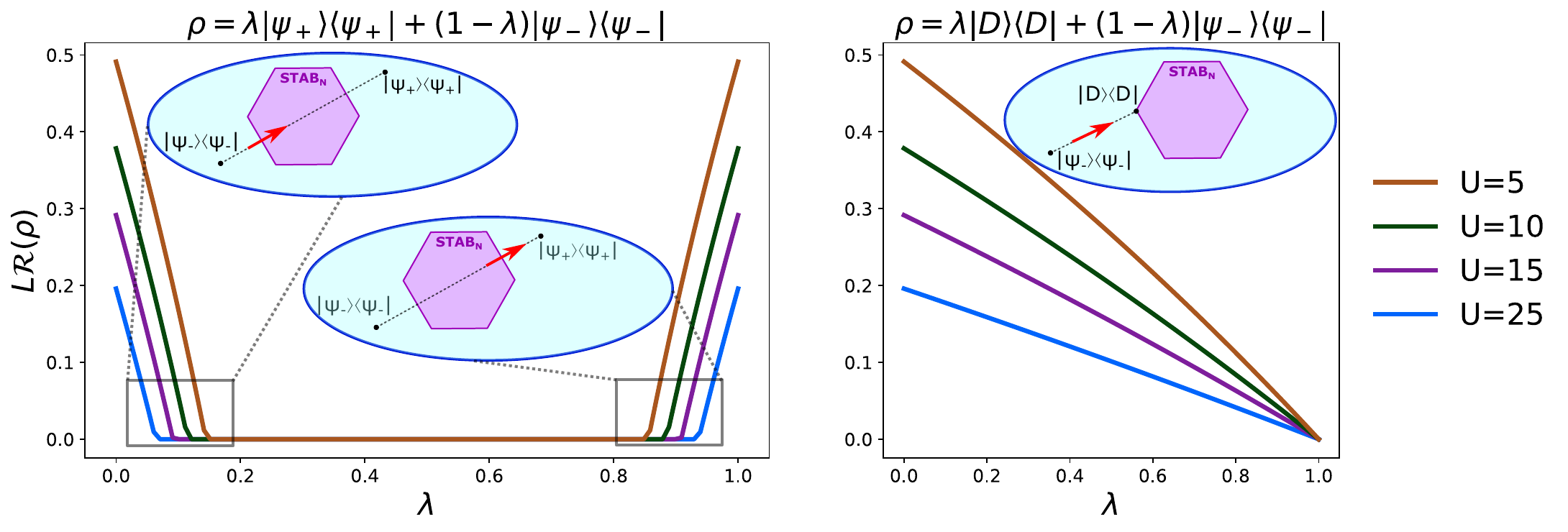}
  \caption{Log-free robustness of magic computed for two different density matrices: on the left panel $\rho=\lambda|\psi_+\rangle\langle\psi_+|+(1-\lambda)|\psi_-\rangle\langle\psi_-|$, while on the right $\rho=\lambda|D\rangle\langle D|+(1-\lambda)|\psi_-\rangle\langle\psi_-|$. We further inserted schematic representations of the linear interpolation between the states $|\psi_-\rangle\langle\psi_-|$, $|\psi_+\rangle\langle\psi_+|$ and $|D\rangle\langle D|$, to show how the different mixing affects the evolution of non-stabilizerness along the path.
  Notice that in the right plot $\LR$ vanishes only when $\lambda=1$, indicating that the state is approaching the stabilizer polytope, but enters it only when the density matrix is fully made of stabilizers. Conversely, on the left the mixture contains the two non-stabilizer states $|\psi_-\rangle$ and $|\psi_+\rangle$, and $\LR$ is zero for $\lambda_\mathrm{c}<\lambda<1-\lambda_\mathrm{c}$, due to the fact that for these values of $\lambda$ the density matrix lies within the stabilizer polytope. 
  }
  \label{fig: AppendixB}
\end{figure*}

As no explicit formula exists beyond the realm of
Gaussian systems, for the two-site Hubbard model we are
stuck to brute-force minimization. 
Fortunately, the limited size of the Hilbert space of the problem allows us to numerically tackle \cref{eq: nonlocal magic app}, with an efficient parametrization of the relevant local unitaries.

The results are shown in \cref{fig: AppendixA}. Clearly, there is an equivalence between the nonlocal magic, obtained from the minimization over local basis changes, and the non-stabilizerness of the two-site Hubbard model. The fact that $\mathcal{M}^\mathrm{NL}(|\psi\rangle)=\mathcal{M}_2(|\psi\rangle)$ rigorously shows that in this model non-stabilizerness is purely nonlocal, and local magic, both in the form of non-stabilizerness of the local reduced density matrix and in the amount of magic that can be removed with local operations, is absent. Note that since we work at $T=0$, the dimer is in a pure (ground) state, and the 2-Renyi entropy $\mathcal{M}_2$ is a faithful measure of magic, for the entire system. This is crucially different than computing $\mathcal{M}_2$ directly for a single site, as discussed in details in the main text. Hence, the equivalence demonstrated by \cref{fig: AppendixA} provides an unambiguous proof that local magic is absent, in term of the SRE family of magic monotones.

In the following, we discuss the numerical methods behind the results of \cref{fig: AppendixA}. 

We begin by noticing that for the two-site Hubbard model, the Hilbert space has dimension $2^4$, and each local unitary lives in a two-qubit Hilbert space of dimension $d_I=4$, and can be represented with a $d_I\times d_I$ complex matrix. Following the approach discussed in \cite{nonlocalmagic_gaussian1}, we parametrize each unitary via the map $U_i(\vec{\theta}_i) = \exp{(-i \vec{\theta}_i \cdot \vec{G})}$ ($i=1,2$) where $\vec{G}$ is the array of the 16 generators of the $U(4)$ group, namely the group of two-qubit unitaries, and $\vec{\theta_i}$ is an array of 16 real angles. We then define the cost function
\begin{equation*}
    C(\vec{\theta}_1,\vec{\theta}_2) = \mathcal{M}_2\left( \, \exp{(-i \,\vec{\theta}_1 \cdot \vec{G})} \otimes \exp{(-i \,\vec{\theta}_2 \cdot \vec{G})} |\psi\rangle \, \right),
\end{equation*}
and we feed it to a gradient-based optimization algorithm based on the Broyden–Fletcher–Goldfarb–Shanno (BFGS) method \cite{bfgs}, which updates the angles until a minimum of the cost function is reached. For the results of \cref{fig: AppendixA}, we choose $|\psi\rangle$ as the ground state of the Hubbard dimer at 
half-filling, \cref{eq: gs}.

\section{Mixing and the vanishing of magic}
\label{app:thermal_magic}

In this appendix, we highlight more explicitly the mechanisms behind both the \emph{thermal} and \emph{dynamical} death of magic shown, respectively, in \cref{sec: FiniteT,sec: Dynamics}. As temperature increases for a fixed interaction strength $U/t$, thermal mixing occurs between the energy eigenstates of the system, namely the non-stabilizer states $|\psi_-\rangle$ and $|\psi_+\rangle$ and the stabilizer states $|D\rangle$ and $|0\rangle$. This mixing alters the structure of the density matrix and, at sufficiently high temperatures the thermal state enters the stabilizer polytope, causing the $\LR$ to vanish. Interestingly, this vanishing is not primarily due to the mixing with stabilizer states, which can only bring the non-stabilizer ground state $|\psi_-\rangle$ arbitrarily close to, but not inside, the stabilizer polytope. Rather, the crucial mechanism is the thermal population of the excited non-stabilizer state $|\psi_+\rangle$: when its Boltzmann weight becomes sufficiently large, the resulting mixture of $|\psi_-\rangle$ and $|\psi_+\rangle$ produces a density matrix that lies entirely within the stabilizer polytope, thus eliminating magic.

To further illustrate this effect, we consider two families of linearly interpolated density matrices, and we compute their non-stabilizerness. In the first case, the mixture $\rho = \lambda\,|\psi_+\rangle\langle\psi_+| + (1-\lambda)\,|\psi_-\rangle\langle\psi_-|$ shows that $\LR(\rho)$ vanishes over a finite range of values of $\lambda$, indicating that the linear combination of the two non-stabilizer states lies within the stabilizer polytope for intermediate mixing ratios. In contrast, if we mix the ground state with a stabilizer state $\rho = \lambda\,|D\rangle\langle D| + (1-\lambda)\,|\psi_-\rangle\langle\psi_-|$, the robustness only vanishes continuously when $\lambda=1$, meaning that the state gradually approaches the polytope, but only enters it when the density matrix is fully composed of stabilizers. Our analysis is summarized in \cref{fig: AppendixB}. This comparison highlights that it is specifically the combination of non-stabilizer states, rather than their admixture with stabilizers, that can drive the system into a stabilizer state. Consequently, the evolution of magic with temperature is closely tied to the redistribution of thermal weight among non-stabilizer eigenstates, which ultimately determines the critical temperature above which magic disappears.


\bibliography{References}

\end{document}